\documentstyle[aaspp4]{article}
\begin{document}

\title{Caustic Crossing Microlensing Event by binary MACHOs and Time Scale Bias}
\author{Mareki Honma\altaffilmark{1}}
\affil{Institute of Astronomy, University of Tokyo, Mitaka, 181-8588, Japan}
\authoremail{honma@milano.mtk.nao.ac.jp}
\altaffiltext{1}{Research Fellow of the Japan Society for the Promotion of Science}
\slugcomment{submitted to the Astrophysical Journal Letter}

\begin{abstract}

Caustic crossing microlensing events provide us a unique opportunity to measure the relative proper motion of the lens to the source, and so those caused by binary MACHOs are of great importance for understanding the structure of the Galactic halo and the nature of MACHOs.
The microlensing event 98-SMC-01, occurred in June 1998, is the first event for which the proper motion is ever measured through the caustic crossing, and this event may be caused by binary MACHOs as we argue in this Letter.
Motivated by the possible existence of binary MACHOs, we have performed the Monte Carlo simulations of caustic crossing events by binary MACHOs and investigated the properties and detectability of the events.
Our calculation shows that typical caustic crossing events have the interval between two caustic crossings ($t_{\rm cc}$) of about 5 days.
We argue that with the current strategy of binary event search the proper motions of these typical events are not measurable because of the short time scale.
Therefore the proper motion distribution measured from caustic crossing events suffers significantly from {`}time scale bias{'}, which is a bias toward finding long time scale events and hence slowly moving lenses.
We predict there are two times more short time scale events ($t_{\rm cc}\le 10$ days) than long time scale events ($t_{\rm cc}\ge 10$ days), and propose an hourly monitoring observation instead of the nightly monitoring currently undertaken to detect caustic crossing events by binary MACHOs more efficiently.
\end{abstract}

\section{Introduction}

Hundreds of gravitational microlensing events have already been detected through the monitoring observations of the stars toward the Magellanic Clouds and the Galactic bulge (Alcock et al.1993; 1996; 1997a; 1997b; Aubourg et al.1993; Alard \& Guibert 1997; Udalski et al.1994a).
While the microlensing events toward the Galactic bulge are probably due to the disk stars, a considerable fraction of the events toward the Magellanic Clouds is likely to be caused by massive compact halo objects (hereafter MACHOs).
The MACHO collaboration (Alcock et al.1997b) has estimated that the MACHO mass is about 0.5 $M_\odot$ and suggested that MACHOs are likely to be old white dwarfs.
However, the mass of the MACHOs is strongly dependent of assumed halo models because of the degeneracy of the distance, velocity and mass for each single event.
In fact, the MACHO mass can be as small as brown dwarfs depending on the halo models (Honma \& Kan-ya 1998).
Moreover there is a possibility that these microlensing events occur not in the halo but in the Magellanic Clouds themselves (Sahu 1994; Sahu \& Sahu 1998), or in some unknown structures in front of the Clouds (Zaritsky \& Lin 1997; Zhao 1998), or in the warped disk of the Galaxy (Evans et al.1998).
These uncertainties make the nature of MACHOs unclear.
In order to understand the nature of MACHOs, it is essential to give further constraints on the Galactic halo.

While microlensing events by single lenses cannot break the degeneracy of the distance, velocity and mass, caustic crossing binary microlensing events provide a unique opportunity to measure the proper motion of lenses to source stars, because the shape of the light curve near the caustic is highly sensitive to the ratio of the source size to the Einstein ring size (Schneider and Weiss 1986).
Although several binary microlensing events have already been found (Udalski et al.1994b; Alard et al.1995; Alcock et al.1997b), intensive monitoring observations near the caustic crossing were not performed until recently.
Such a monitoring observation has been made for the first time for the event 98-SMC-01 (Afonso 1998; Albrow et al.1998; Alcock el al.1998), which occurred in June 1998.
From the light curve near the caustic, they have obtained the proper motion of $1\sim 2$ km/s/kpc, which suggests that the lens is likely to be binary stars in the SMC.

However, as will be shown in the next section, the possibility that the lens of the event 98-SMC-01 is in the halo cannot be ruled out when one considers the event rate for the halo and the SMC self-lensing.
If the event 98-SMC-01 is due to binary MACHOs, this implies that there exist plenty of binary MACHOs in the Galactic halo, as one out of two microlensing events so far detected towards the SMC shows the feature of the binary microlensing.
Therefore, it is interesting to investigate the properties of caustic crossing events by binary MACHO.
In particular, it is of significant importance to study an observational bias that comes from the current strategy of caustic crossing event search.
As is the case for 98-SMC-01, caustic crossing events are likely to be realized after the first caustic crossing: generally it is difficult to detect the first caustic crossing with usual nightly observations, but it is relatively easier to recognize the rise of the source brightness when the source star comes into the caustic.
In fact 98-SMC-01 was first recognized as a possible binary event three days after the first caustic crossing (Alcock et al.1998), and then intensive monitoring observations of the second caustic crossing, which occurred about 12 days after the first caustic crossing, were performed (Afonso et al.1998; Albrow et al.1998; Alcock et al.1998).
This strategy for finding caustic crossing events is efficient for long time scale events, but is likely to miss events with short time scales (e.g., $t_{\rm cc}$ less than 5 days).
Thus the current strategy probably introduces {`}time scale bias{'}, which is a bias toward finding long time scale events and hence slowly moving lenses.
Therefore, for further search and observations of caustic crossing events by binary MACHOs, it is important to understand the effect of the time scale bias.

The plan of this Letter is as follows : in section 2 we describe the models we use in this Letter. We also present the differential event rate with respect to the proper motion for halo and SMC models, and show that 98-SMC-01 could be an event by binary MACHOs in the halo.
In Section 3, we present the Monte Carlo simulation of the caustic crossing events and investigate the effect of the time scale bias.
Discussions on the future observations of binary MACHOs will be made in Section 4.

\section{Halo Models and Differential Event Rate Distribution}

In this section we describe the models for the halo and SMC self-lensing and calculate differential event rate with respect to the proper motion $\mu$.
We consider following three models: the standard halo case, the Plummer halo case which is used by Honma \& Kan-ya (1998), and the SMC self-lensing case.
We use two different halo models because recent investigations imply that the Galactic rotation curve may deviate from that of the standard halo model (Honma \& Sofue 1996; 1997; Olling \& Merrifield 1998).

What we can calculate is the single event rate assuming that all lensing objects are single MACHOs or stars.
The binary event rate cannot be obtained easily because it is not only a function of halo model parameters, but also a function of binary fraction and separation distribution.
Since we have no information of these binary properties of MACHOs or SMC stars, here we assume that there are no difference in binary fraction or separation distributions in the SMC and halos, and simply assume that the rate for binary events are just proportional to the single event rate.

For the standard halo case, we use the density and velocity distribution of Griest (1991) with the core radius $a$ of 5 kpc and the rotation velocity in the halo of 220 km/s.
The local dark matter density is assumed to be $1.0\times 10^{-2} M_\odot$/pc$^2$.
The MACHO mass $m$ and the halo MACHO fraction $f$ are set to be $0.5M_\odot$ and $0.5$ following Alcock et al.(1997b).
We assume the distance to the Galactic center of 7.5 kpc and the distance to the SMC of 60 kpc. 
The standard halo model predicts the total event rate $\Gamma$ of $1.6\times 10^{-6}$ per year and the optical depth $\tau$ of $3.8 \times 10^{-7}$ toward the SMC.
For the Plummer halo case, we set the core radius of $12$ kpc and the total mass of $1.1\times 10^{11}M_\odot$ (Honma \& Kan-ya 1998; hereafter HK halo model).
Since this model yields the MACHO mass as small as brown dwarfs, we assume the MACHO mass of $0.08M_\odot$ for the HK halo case.
This model gives $\Gamma=1.8\times 10^{-6}$ and $\tau=3.3\times 10^{-7}$, which are comparable to those of the standard halo models.
For the two halo cases, we neglect the proper motion of the source because the source motion changes only slightly the distribution of the relative proper motion (Albrow et al.1998), but the source proper motion is taken into account for the SMC self-lensing case.
As for the SMC self lensing case, though there are some suggestions that the SMC may be elongated along the line of sight or it may have some substructures (e.g.,Mathewson et al.1986; Sahu \& Sahu 1998), we model the SMC with a Plummer model with the core  radius of 5 kpc and the total mass of $3\times 10^9M_\odot$ (for SMC parameters, see, e.g., Gardiner \& Noguchi 1996; Palanque-Delabrouille et al.1998).
Note that the central velocity dispersion of this model is 21 km/s, which agrees well with the observation (Hatzidimitriou et al.1997).
Only for the SMC case we assume that the source star is at the distance of 65 kpc to make the lensing probability relatively higher.
This model for SMC self-lensing gives $\Gamma=1.2\times 10^{-7}$ and $\tau=1.1\times 10^{-7}$.
Note that the event rate for SMC self-lensing is smaller than those for the halos by an order of magnitude.
If the SMC is elongated along the line of sight or if it has substructures, the optical depth may be increased by a factor of about 3 (Palanque-Delabrouille 1998; Sahu \& Sahu 1998), but not by an order of magnitude.

In figure 1 we show the differential event rate with respect to the proper motion $\mu$ as well as the possible range of the proper motion of 98-SMC-01 event.
Albrow et al.(1998) found two possible solutions for this event, corresponding to $\mu=1.26$ or 2.00 km/s/kpc.
On the other hand, Alcock et al.(1998) obtained $\mu=1.4$ km/s/kpc.
At present it is unclear which of the models best represent the event, and so we assume that the true proper motion of 98-SMC-01 lies between $1.0$ and $2.2$ km/s/kpc including the error.
As seen in figure 1, the event rates in the observed range of the proper motion are similar to each other for the HK halo case and the SMC self-lensing case, while the event rate for the standard halo case is moderately smaller.
The ratio of the event rate for $\mu$ between 1.0 and 2.2 for the HK case halo to that for SMC self-lensing case is
\begin{equation}
\frac{\Gamma_{\rm HK} (1.0 \le\mu\le 2.2)}{\Gamma_{\rm SMC} (1.0 \le\mu\le 2.2)} = 1.7,
\end{equation}
and the ratio of that for the standard halo case to that for SMC case is
\begin{equation}
\frac{\Gamma_{\rm ST} (1.0 \le\mu\le 2.2)}{\Gamma_{\rm SMC} (1.0 \le\mu\le 2.2)} = 0.51.
\end{equation}
Therefore, as long as the ratio of binary event rate to the single event rate does not vary significantly between the halos and the SMC, the probability that the 98-SMC-01 is due to the halo binary MACHOs is comparable to that of the SMC self-lensing.
In particular, the HK halo model gives higher probability for the halo binary MACHOs than that for the SMC self-lensing.
Hence, at this stage, whether the lens of the event 98-SMC-01 is in the halo or in the SMC is still unclear.

\section{Properties of Caustic Crossing Event by binary MACHOs} 

In the previous section we have seen that the possibility of a halo binary is as high as an SMC binary for the event 98-SMC-01.
If this event is caused by binary MACHOs, it may indicate that there are plenty of binary MACHOs in the Galactic halo, and thus it is worth studying the properties of the caustic crossing events caused by binary MACHOs.
In particular, as described in the Introduction, there exists an observation bias toward finding long time scale events.
In order to investigate the effect of the bias, we have to know the distribution of the interval between two caustic crossings, $t_{\rm cc}$.
Since $t_{\rm cc}$ depends on many parameters such as the shape of caustic, impact parameter of MACHOs, the distance and velocity of MACHOs and so on, $t_{\rm cc}$ cannot be expressed analytically, and hence we use the Monte Carlo simulations to calculate the distribution of $t_{\rm cc}$ for caustic crossing events.

We have simulated about $3 \times 10^{4}$ caustic crossing events for the standard halo case, the HK halo case, and SMC self-lensing case.
The distance and velocity of the lensing objects were generated according to the density and velocity distribution of the halo or SMC models described in the previous section.
The binary separation distribution was assumed to be in the form of $dN/da\propto a^{-1}$, which corresponds to the distribution for the main sequence stars (Abt 1983).
In each model the mass of the MACHOs and SMC lensing stars was assumed to be unique: we set $m=0.5M_\odot$ for the standard halo, $0.08M_\odot$ for the HK halo, and $0.2M_\odot$ for the SMC. 
The projection of the binary system onto the sky plane was assumed to be at random, and the caustics were calculated by solving the lensing equation numerically.
For some particular cases, a source can cross the caustic more than twice.
Here we neglect these multiple caustic crossing events and concentrate on the events with two caustic crossings.
According to our calculation, the fraction of multiple caustic crossing events is about 8\% out of all caustic crossing events.

Figure 2 shows the integrated probability $P[t_{\rm cc}\le t]$ , which is the probability for finding an event with $t_{\rm cc}$ less than $t$.
For the standard and HK halo cases, $P[t_{\rm cc}\le t]$ becomes 0.5 at $t=6.6$ and 5.3 days, respectively.
For comparison, we note that typical Einstein ring crossing time $t_{\rm E}(=R_E/v_\perp)$ for single lens is about 40$\sim$50 days.
This implies that the path of the source inside the caustic is much smaller than the size of the Einstein ring.
This can be explained partly by the fact that the size of caustics is usually much smaller than the Einstein ring size, and partly by the fact that the cuspy structure of caustics allow a large number of events for which the path in the caustic is very short.

The short time scale for binary MACHOs implies that proper motions are not measurable for typical caustic crossing events by binary MACHOs.
In the current strategy, caustic crossing events are recognized a few days after the first caustic crossing based on the rise of the source brightness, and further monitoring for some more days is required to predict the precise date of the second caustic crossing.
Therefore, $t_{\rm cc}$ of a week or more is required for intensive monitoring of the second caustic crossing.
For simplicity, here we assume that we cannot measure the proper motion of caustic crossing events if $t_{\rm cc}$ is less than 10 days (in fact 98-SMC-01 has $t_{\rm cc}$ of $\sim 12$ days).
The fraction of the caustic crossing events with $t_{\rm cc}$ less than 10 days is 0.6 to 0.7 for the halo cases while it is about 0.3 for the SMC self-lensing case.
Therefore, for the halo binary cases, the fraction of the events for which proper motions are not measurable is much more than the events for which proper motions are measurable.

To see the effect of the bias more clearly we show in figure 3 the average of the interval between two caustic crossings, $\bar{t}_{\rm cc}$, for each proper motion bin.
In each bin about 60$\sim$70 \% of events have $t_{\rm cc}$ less than $\bar{t}_{\rm cc}$, but there also exist some events with $t_{\rm cc}$ over 50 days, which make the dispersion in each bin quite large.
Nevertheless, $\bar{t}_{\rm cc}$ is generally decreasing with increasing the proper motion $\mu$, as is expected.
$\bar{t}_{\rm cc}$ becomes less than 10 days beyond $\log \mu=1.4$ for the standard halo case, and $\log \mu=1.0$ for the HK halo case.
From figure 1, the distribution of proper motion is peaked at $\log \mu=1.3$ for the standard halo and $\log \mu=1.1$ for the HK halo model.
Therefore, for both cases, proper motions are not measurable for the caustic crossing events that have most probable value of $\mu$, and hence we conclude that the time scale bias affect significantly the distribution of $\mu$ measured from caustic crossing events.
The bias would explain why the proper motion of 98-SMC-01 appears at the lower end of the event rate distribution for the halos in figure 1.
As far as events with $t_{\rm cc}$ between 10 and 15 days we have simulated are concerned, the fraction of events that have $\mu$ between 1.0 and 2.2 km/s/kpc is 5 \% for the halo cases, and so the possibility of binary MACHOs cannot be ruled out statistically.

\section{Proposal for high frequency monitoring}

From figure 2 one can see that most of SMC self-lensing events have $t_{\rm cc}$ larger than 10 days while many of halo binary events have $t_{\rm cc}$ shorter than 10 days.
Therefore, it is interesting to search for short time scale events, with which one can discriminate whether binary lenses are likely to exist in the halo or the Magellanic Clouds.
Here we define the ratio $R$ of the number of events with $t_{\rm cc}$ shorter than 10 days to the number of events with $t_{\rm cc}$ longer than 10 days, namely,
\begin{equation}
R=\frac{P[t_{\rm cc} \le 10\: {\rm days}]}{P[t_{\rm cc} \ge 10\; {\rm days}]}.
\end{equation}
From figure 2 we obtain $R_{\rm ST}=1.5$ for the standard halo case, $R_{\rm HK}=1.9$ for the HK halo case, and $R_{\rm SMC}=0.47$ for the SMC self-lensing events.
Therefore, if the binary lens of 98-SMC-01 is in the SMC itself as claimed by previous studies, there should have been no other caustic crossing events with shorter time scale.
On the other hand, if 98-SMC-01 is caused by the binary MACHOs, one or two more caustic crossing events should have already occurred other than 98-SMC-01.
These events were probably missed for intensive monitoring of the caustic crossing because of the short time scale, but the variation of brightness may have been recorded in the usual nightly observations.
Although the number of the data points between two caustic crossings is small for these events, they may be detectable as is the case for the event DUO-2, which was found toward the Galactic bulge (Alard et al.1995).
Therefore, searching carefully for short time scale events in the data could provide some useful information on the existence of binary MACHOs.

To measure the proper motion for these short time scale events, a high frequency monitoring observation, {`}hourly{'} observation rather than {`}nightly{'} observation, is necessary.
The Japan-New Zealand collaboration MOA, which observes the source stars several times per night, may be one of the most suitable projects for this kind of study.
Also important is the international collaboration of monitoring groups so that the caustic crossing events with short time scale can be monitored all day and night.
If short time scale caustic crossing events caused by binary MACHOs are found and their proper motions are measured, they are of great importance for discriminating the halo models: while the standard halo model gives $\bar{\mu}$ of 38 km/s/kpc, the HK halo model predicts $\bar{\mu}$ of 26 km/s/kpc.
Therefore, detecting a lager number of binary events and measuring their proper motions is a promising approach for constraining the halo model.
From these caustic crossing events, we can learn not only about the kinematics of the Galactic halo, but also about the nature of MACHOs, because the MACHO mass strongly depends on the halo models.

\acknowledgments

We are grateful to Y. Sofue, Y. Kan-ya and H. Kamaya for helpful discussions, and T. Sato for his continuous encouragement.
We also thank the Japan Society for the Promotion of Science for their financial support.

\clearpage

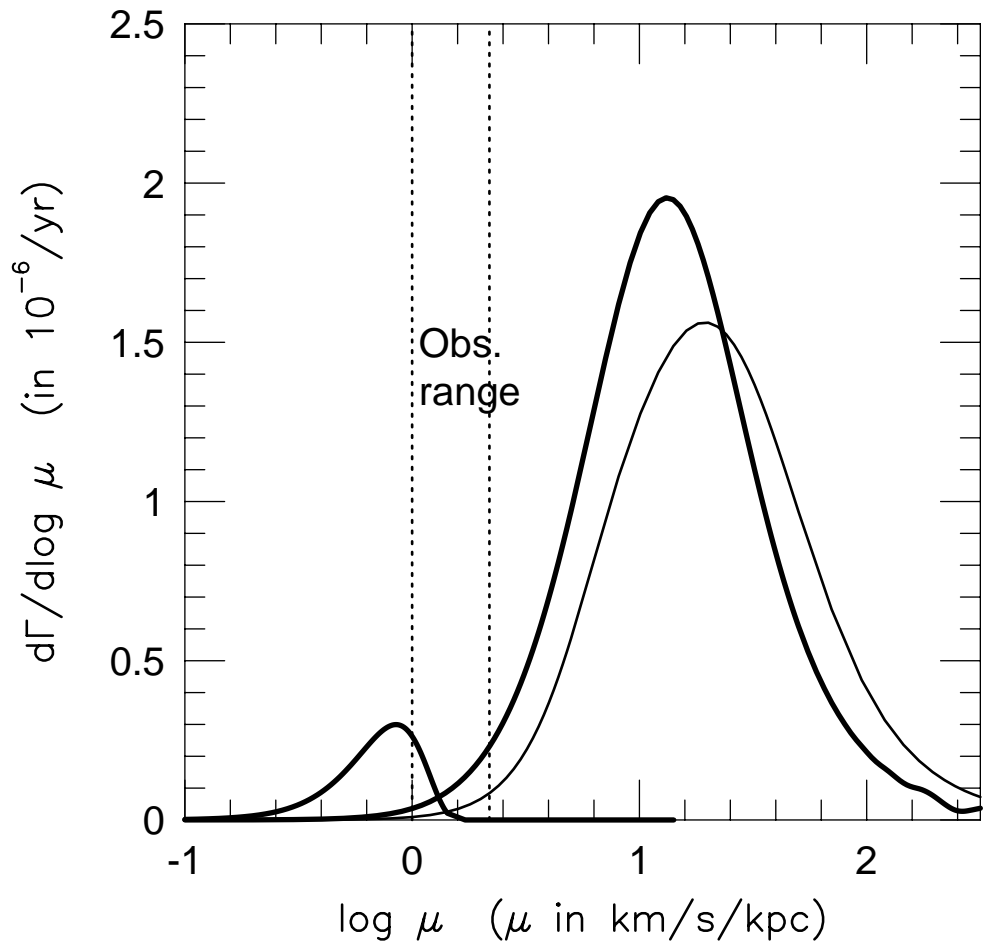
\begin{figure}
\caption{Differential event rate with respect to the relative proper motion $\mu$ for the standard halo case (thin line), the HK halo case (thick line, right) and the SMC self-sensing case (thick line, left).
The probable range of the proper motion for 98-SMC-01 is also shown.
}
\end{figure}

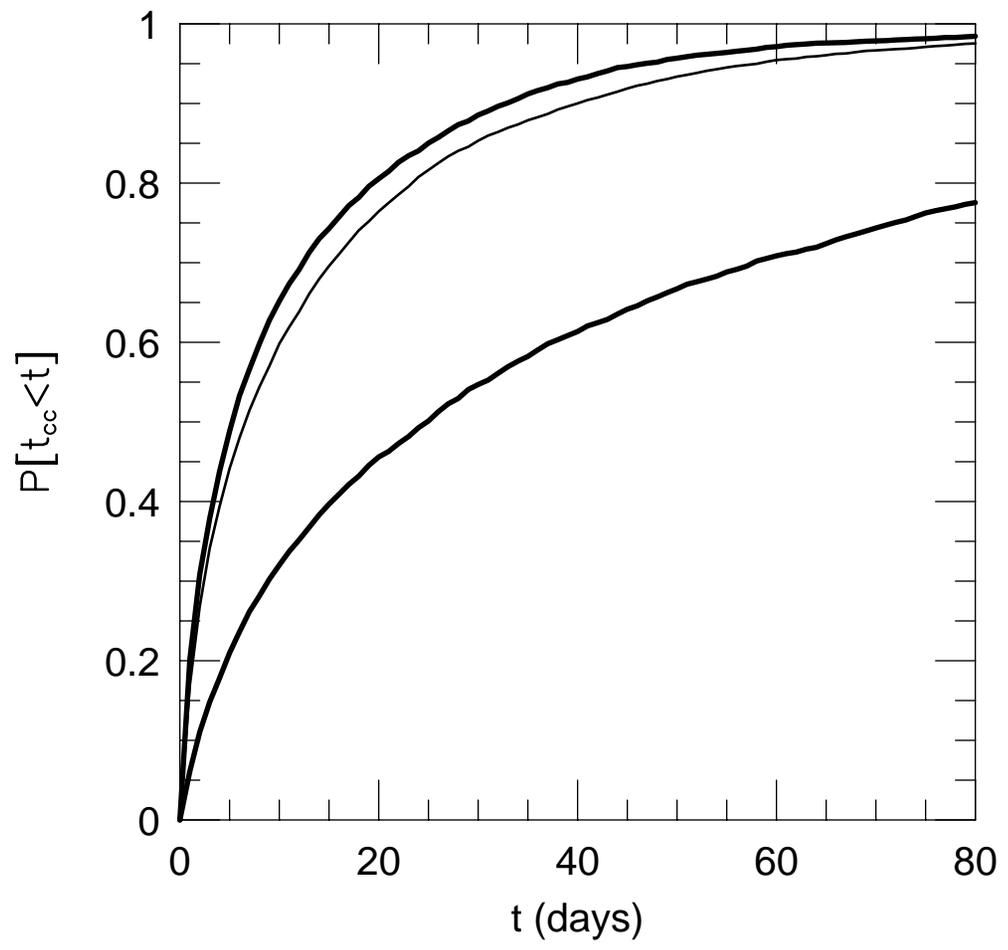
\begin{figure}
\caption{Integrated probability for finding an event with $t_{\rm cc}$ less than $t$.
Thin line is for the standard halo case, upper thick line is for the HK halo case, and lower thick line is for the SMC self-lensing case.
}
\end{figure}

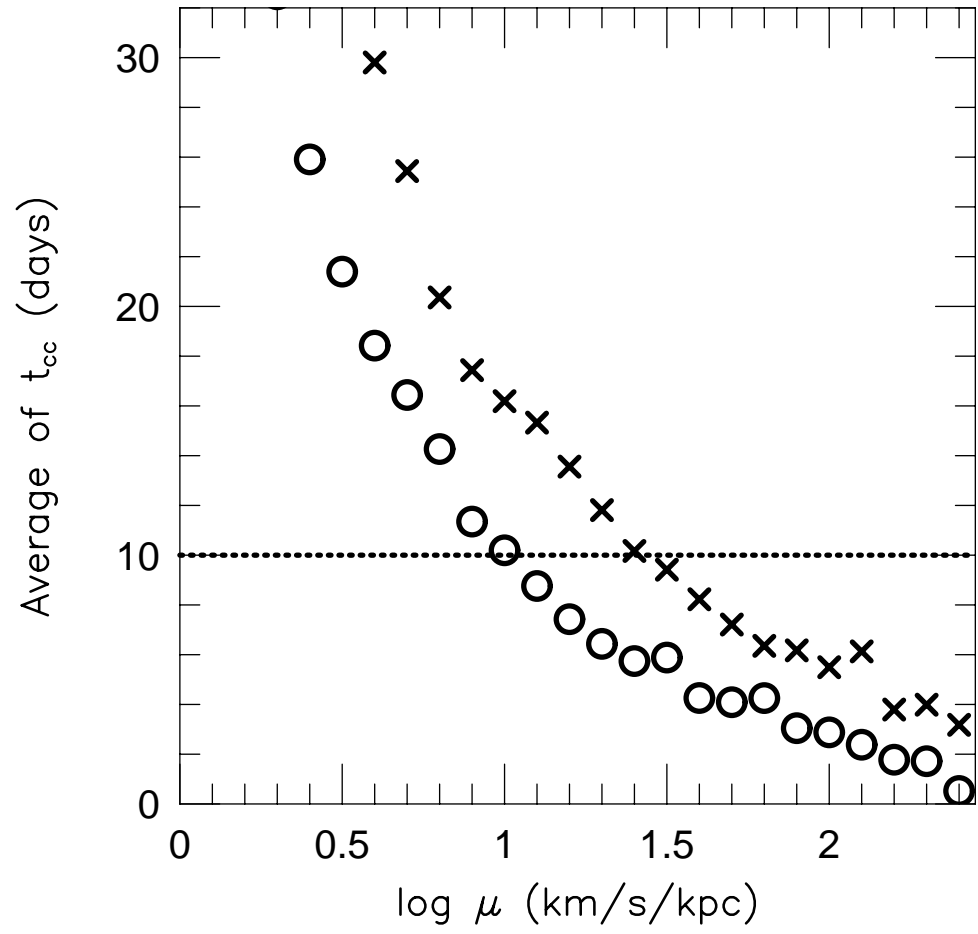
\begin{figure}
\caption{Average of $t_{\rm cc}$ against $\log \mu$.
Circles are for the HK halo case, and crosses for the standard halo cases.
The horizontal dotted line corresponds to $\bar{t}_{\rm cc}=10$ days, below which the proper motions are not measurable in the current observation.
}
\end{figure}

\end{document}